\documentclass[3p]{elsarticle}

\usepackage[breaklinks=true]{hyperref}
\usepackage[anythingbreaks]{breakurl}
\usepackage{amsthm}
\usepackage{amsmath}
\usepackage{amsfonts}
\usepackage{amssymb}
\usepackage{amsfonts}
\usepackage{epsfig}
\usepackage{bbm}
\usepackage{todonotes}
\usepackage{soul}
\usepackage{nicefrac}
\usepackage{booktabs}
\usepackage{siunitx}
\usepackage[capitalise]{cleveref}
\usepackage[super]{nth}
\usepackage{float}
\usepackage{caption}
\usepackage{subcaption}

\usetikzlibrary{arrows.meta, 
	bending, 
	calc, chains, 
	decorations.pathmorphing,
	positioning
}

\usepackage{pgfplots}

\usepgfplotslibrary{fillbetween}
\usetikzlibrary{intersections}

\newcommand{\deltat}{\tilde{\delta}}

\newcommand{\tens}[1]{\boldsymbol{\mathrm{#1}}}

\journal{Journal of \LaTeX\ Templates}
\definecolor{rwth}   {RGB}{  0  84 159}
\definecolor{rwth-75}{RGB}{ 64 127 183}
\definecolor{rwth-50}{RGB}{142 186 229}
\definecolor{rwth-25}{RGB}{199 221 242}
\definecolor{rwth-10}{RGB}{232 241 250}

\definecolor{black}   {RGB}{  0   0   0}
\definecolor{black-75}{RGB}{100 101 103}
\definecolor{black-50}{RGB}{156 158 159}
\definecolor{black-25}{RGB}{207 209 210}
\definecolor{black-10}{RGB}{236 237 237}

\definecolor{magenta}   {RGB}{227   0 102}
\definecolor{magenta-75}{RGB}{233  96 136}
\definecolor{magenta-50}{RGB}{241 158 177}
\definecolor{magenta-25}{RGB}{249 210 218}
\definecolor{magenta-10}{RGB}{253 238 240}

\definecolor{yellow}   {RGB}{255 237   0}
\definecolor{yellow-75}{RGB}{255 240  85}
\definecolor{yellow-50}{RGB}{255 245 155}
\definecolor{yellow-25}{RGB}{255 250 209}
\definecolor{yellow-10}{RGB}{255 253 238}

\definecolor{petrol}   {RGB}{  0  97 101}
\definecolor{petrol-75}{RGB}{ 45 127 131}
\definecolor{petrol-50}{RGB}{125 164 167}
\definecolor{petrol-25}{RGB}{191 208 209}
\definecolor{petrol-10}{RGB}{230 236 236}

\definecolor{turkis}   {RGB}{  0 152 161}
\definecolor{turkis-75}{RGB}{  0 177 183}
\definecolor{turkis-50}{RGB}{137 204 207}
\definecolor{turkis-25}{RGB}{202 231 231}
\definecolor{turkis-10}{RGB}{235 246 246}

\definecolor{grun}   {RGB}{ 87 171  39}
\definecolor{grun-75}{RGB}{141 192  96}
\definecolor{grun-50}{RGB}{184 214 152}
\definecolor{grun-25}{RGB}{221 235 206}
\definecolor{grun-10}{RGB}{242 247 236}

\definecolor{maigrun}   {RGB}{189 205   0}
\definecolor{maigrun-75}{RGB}{208 217  92}
\definecolor{maigrun-50}{RGB}{224 230 154}
\definecolor{maigrun-25}{RGB}{240 243 208}
\definecolor{maigrun-10}{RGB}{249 250 237}

\definecolor{orange}   {RGB}{246 168   0}
\definecolor{orange-75}{RGB}{250 190  80}
\definecolor{orange-50}{RGB}{253 212 143}
\definecolor{orange-25}{RGB}{254 234 201}
\definecolor{orange-10}{RGB}{255 247 234}

\definecolor{rot}   {RGB}{204   7  30}
\definecolor{rot-75}{RGB}{216  92  65}
\definecolor{rot-50}{RGB}{230 150 121}
\definecolor{rot-25}{RGB}{243 205 187}
\definecolor{rot-10}{RGB}{250 235 227}

\definecolor{bordeaux}   {RGB}{161  16  53}
\definecolor{bordeaux-75}{RGB}{182  82  86}
\definecolor{bordeaux-50}{RGB}{205 139 135}
\definecolor{bordeaux-25}{RGB}{229 197 192}
\definecolor{bordeaux-10}{RGB}{245 232 229}

\definecolor{violett}   {RGB}{ 97  33  88}
\definecolor{violett-75}{RGB}{131  78 117}
\definecolor{violett-50}{RGB}{168 133 158}
\definecolor{violett-25}{RGB}{210 192 205}
\definecolor{violett-10}{RGB}{237 229 234}

\definecolor{lila}   {RGB}{122 111 172}
\definecolor{lila-75}{RGB}{155 145 193}
\definecolor{lila-50}{RGB}{188 181 215}
\definecolor{lila-25}{RGB}{222 218 235}
\definecolor{lila-10}{RGB}{242 240 247}

\begin{document}

\begin{frontmatter}
	\title{Discovering Asymptotic Expansions Using Symbolic Regression}
	
	\author[km]{Rasul Abdusalamov\corref{mycorrespondingauthor}}
		\cortext[mycorrespondingauthor]{Corresponding author}
		\ead{abdusalamov@km.rwth-aachen.de}
	\author[UK]{Julius Kaplunov}
	\author[km]{Mikhail Itskov}

	\address[km]{Department of Continuum Mechanics, RWTH Aachen University, Germany}
	\address[UK]{School of Computer Science and Mathematics, Keele University, United Kingdom}

	\begin{abstract}		
	Recently, symbolic regression (SR) has demonstrated its efficiency for discovering basic governing relations in physical systems. A major impact can be potentially achieved by coupling symbolic regression with asymptotic methodology. The main advantage of asymptotic approach involves the robust approximation to the sought for solution bringing a clear idea of the effect of problem parameters. However, the analytic derivation of the asymptotic series is often highly nontrivial especially, when the exact solution is not available. \\
	In this paper, we adapt SR methodology to discover asymptotic series. As an illustration we consider three problem in mechanics, including two-mass collision, viscoelastic behavior of a Kelvin-Voigt solid and propagation of Rayleigh-Lamb waves. The training data is generated from the explicit exact solutions of these problems. The obtained SR results are compared to the benchmark asymptotic expansions of the above mentioned exact solutions. Both convergent and divergent asymptotic series are considered. A good agreement between SR expansions and analytical results is observed. It is demonstrated that the proposed approach can be used to identify material parameters, e.g. Poisson's ratio, and has high prospects for utilizing experimental and numerical data.
	\end{abstract}

\end{frontmatter}

\section{Introduction}
\label{sec:Introduction}
Nowadays with pretense of data, the field of machine learning (ML) gains a major role in scientific research. Recent ML applications range from reducing measurement errors in quantum computations \cite{Seif_2018} to the acceleration of fluid dynamics simulations \cite{kochkov2021machine}. At the same time ML based algorithms have certain disadvantages. In particular, ML suffers from a lack of interpretability, since the algorithms employed are "black box" models. It is often difficult to gain qualitative insights into such models and to fully interpret their behavior. In addition, ML training is usually expensive with respect to computational costs and other resources. Furthermore, biased data may result in inaccurate predictions. Finally, ML algorithms may overfit on a limited data set leading to a poor generalization. \\
In recent years, symbolic regression (SR) demonstrated substantial potential in addressing some of the above mentioned disadvantages of ML algorithms \cite{Orzechowski2018, wang_wagner_rondinelli_2019}. The key idea of SR, as described by Augusto et. al. \cite{AugustoSR2000}, is to establish the structure of an appropriate mathematical model aimed at describing given data. This is achieved by specifying a pool of functions, operations and inputs forming a solution space. The main advantage of this approach is that no a priori assumptions need to be made about the sought for structure of the model. Koza \cite{koza1994genetic} introduced an evolutionary computational technique, known as genetic programming, for searching the solution  space. A variety of libraries and frameworks have been developed since then, e.g. see \cite{la2021contemporary} reporting on their performance. At the moment SR is implemented in many areas, including discovering the governing equations for an elastic Timoschenko beam \cite{AITimoshenko}, reconstructing orbital anomalies \cite{manzi2020orbital}, accelerating the discovery of novel catalysts \cite{weng2020simple} as well as investigating dynamic systems \cite{gaucel2014learning} just to mention a few. \\
The implementation of the SR technique may greatly benefit from preliminary physical analysis of the tackled problem, including a definition of problem parameters and scaling laws. This is why the asymptotic analysis has a substantial potential in this field, e.g. see \cite{andrianov2002asymptotology}. Asymptotic analysis is a powerful method for simplifying complex relationships to estimate their limiting behavior. It is hardly possible to make here a proper account of the current state of the art in general area of asymptotic methods. Here, we restrict ourselves by mentioning several influential books on the subject, e.g. \cite{bender1999advanced, kevorkian2013perturbation, Fedoryuk, naife1984introduction, copson2004asymptotic, de1981asymptotic, simmonds1998first, bauer2015asymptotic} and references therein. At the same time, asymptotic routines can also get a new powerful impulse from adapting SR. In particular, the calculation of higher order terms may be facilitated in asymptotic expansion even when the exact analytic solution is known but cumbersome. Moreover, SR may be instrumental when the solution is found by a numerical procedure, e.g. using FEM software. In addition, SR appears to be able to extract an asymptotic series from experimental data. Thus, the combination of these two rather different approaches is highly promising for making a substantial impact on the modern research methodology. \\
In this paper we make an initial effort to apply SR to basics problems in mechanics. Each of them has an explicit exact solution and also allows asymptotic expansions in terms of small or large problem parameters. The exact solutions are used for generating artificial training data to discover SR approximations. However, due to the physical origin of the considered examples, the training data can be equally taken from experimental measurements. The benchmark asymptotic series help to evaluate the accuracy of the obtained SR results. \\
The paper is organized as follows. \autoref{sec:Methodology} is concerned with a general introduction into symbolic regression mentioning the prospect for asymptotic series. The simplest example of a two-mass collision problem is considered in \autoref{sec:InitialImpactProblem}. Despite its simplicity this problem demonstrates three different types of asymptotic behavior. All of them are given by convergent series. An example of a divergent asymptotic series is presented in \autoref{sec:KelvinVoigtModel}, dealing with a viscoelastic Kelvin-Voigt model. Finally, bending wave propagation in an elastic layer is analyzed in \autoref{sec:ElasticLayer}. The previous asymptotic consideration for Rayleigh-Lamb waves, e.g. see \cite{kaplunov1998dynamics, gol1990asymptotic, goldenveizer1993timoshenko} are adapted for establishing an SR series. In addition, the obtained SR results are applied for the evaluation of Poisson's ratio. A conclusion and outlook are given in \autoref{sec:Conclusions}.

\section{Theoretical Background}
\label{sec:Methodology}

\begin{figure}[ht!]
	\centering
	\includegraphics[width=0.8\textwidth]{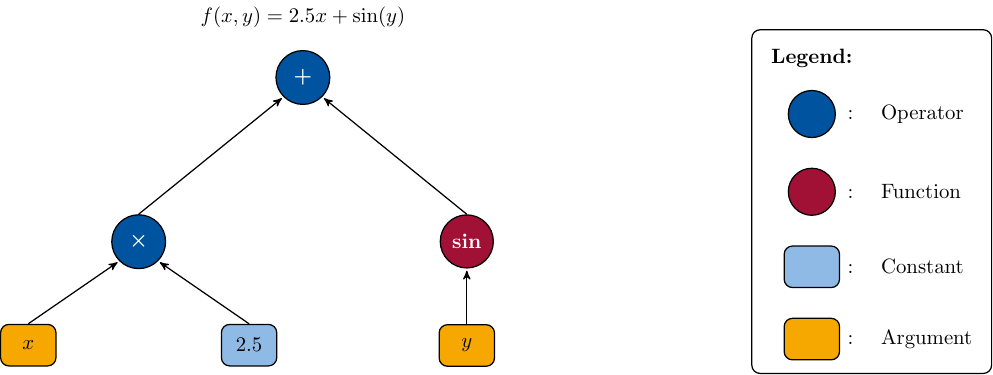}
	\caption{A graphical tree of an algebraic expression.}
	\label{fig:TreeExample}
\end{figure}

\begin{figure} [ht!]
	\centering
	\includegraphics[width=1\textwidth]{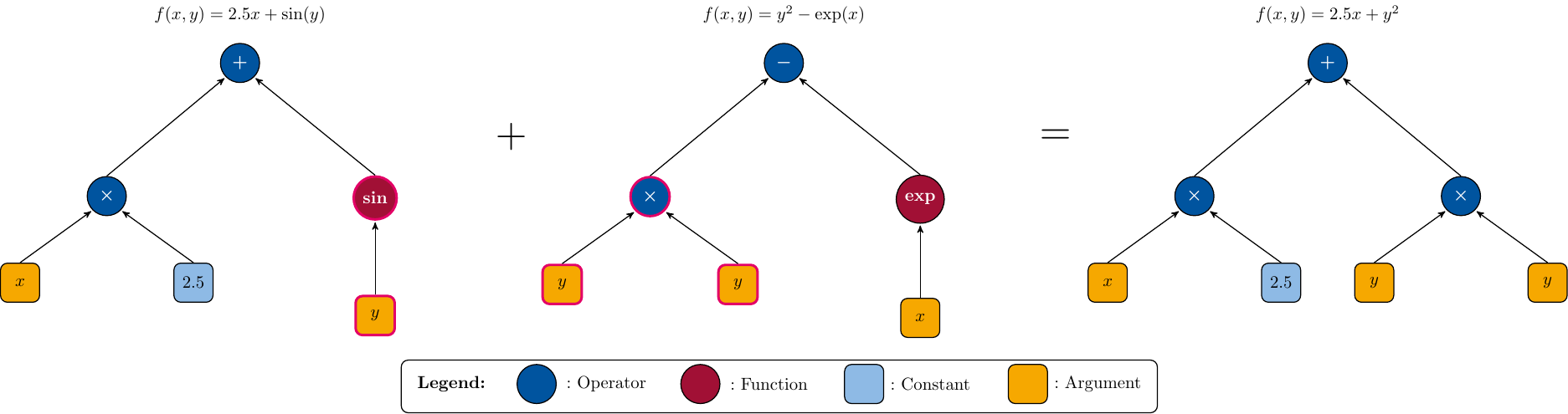}
	\caption{Example of a crossover mutation.}
	\label{fig:CrossoverExample}
\end{figure}

\begin{figure} [ht!]
	\centering
	\includegraphics[width=.9\textwidth]{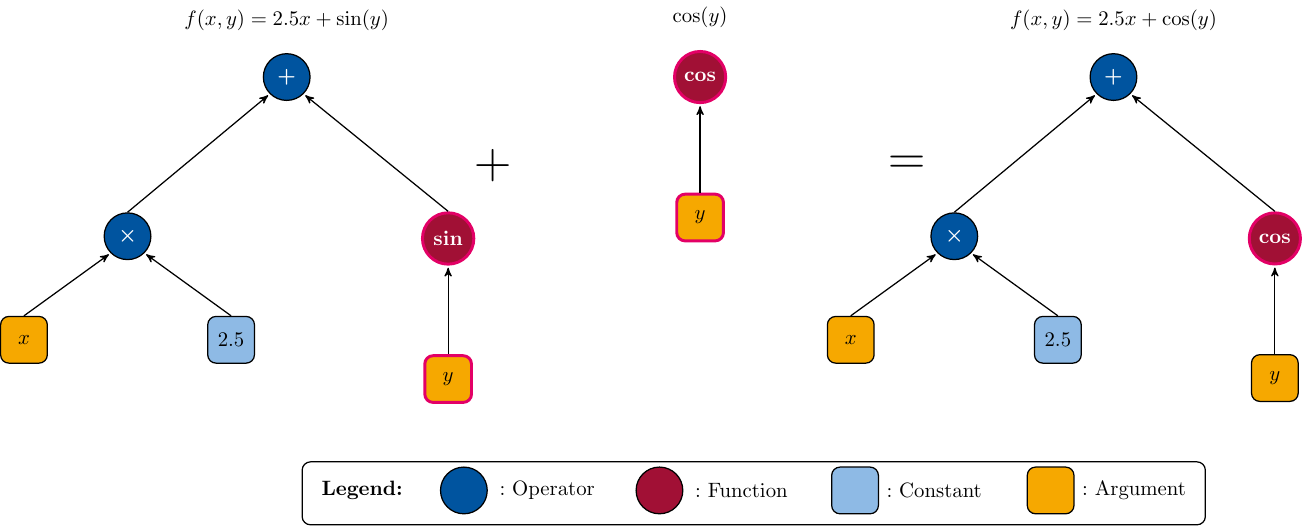}
	\caption{Example of a subtree mutation.}
	\label{fig:SubtreeExample}
\end{figure}

\begin{figure}[ht!]
	\centering
	\includegraphics[width=.8\textwidth]{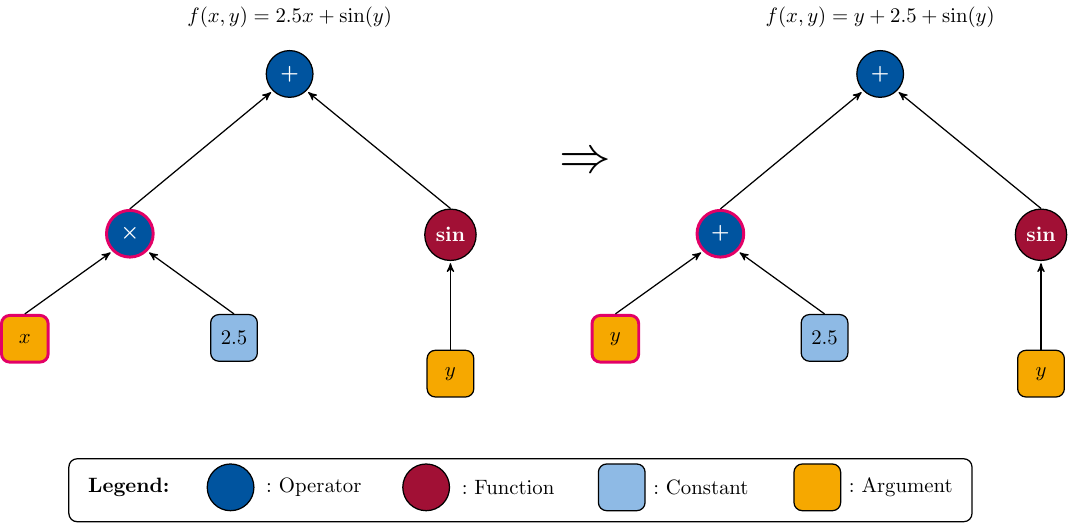}
	\caption{Example of a point mutation.}
	\label{fig:PointExample}
\end{figure}

\begin{figure}[ht!]
	\centering
	\includegraphics[width=.8\textwidth]{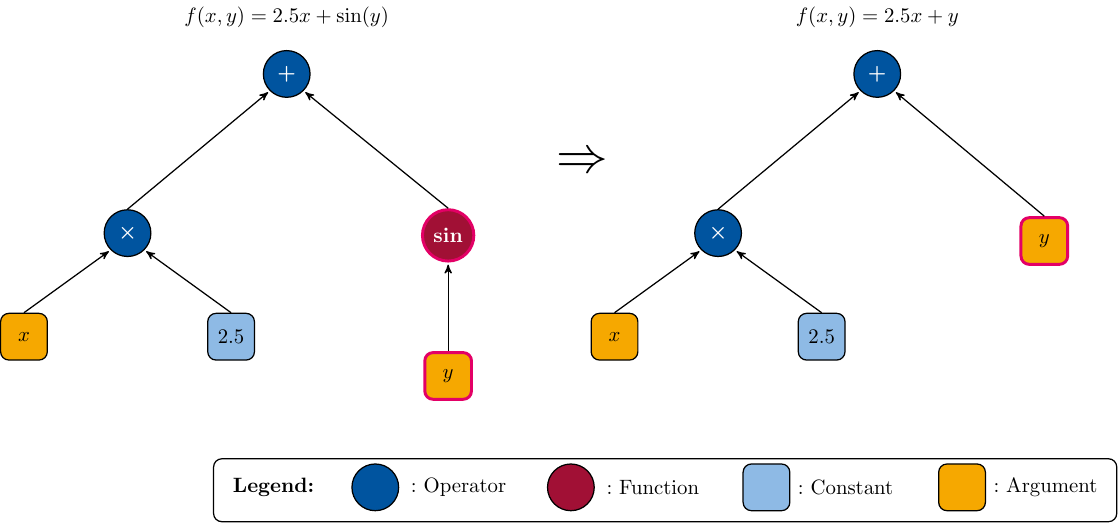}
	\caption{Example of a hoist mutation.}
	\label{fig:HoistExample}
\end{figure}

In the traditional sense, regression is a statistical technique that identifies the relationship between a single dependent variable and one or more independent variables. Typically, an a priori model structure, such as a linear model, is used to determine the best fit for a given set of data. Predefined parameters of the model are optimised. In the case of symbolic regression, no assumptions are made about the model structure or type. SR finds the ideal structure and the relationship between the independent variables and the dependent variable \cite{AugustoSR2000}. The result is an algebraic expression that optimally describes the given data set. Typically, such expressions can be described in the form of graphical trees that place operations, constants and inputs in hierarchical relationships (see \autoref{fig:TreeExample}).
To find an optimal formulation, most symbolic regression frameworks use genetic programming (GP), an evolutionary computation algorithm. This approach was originally introduced by Koza \cite{koza1994genetic} and utilizes a hierarchical function definition to automatically and dynamically identify potentially candidates. By generating a population of possible solutions and evolving them over a specified number of generations a large search space can be searched in an efficient way. So far SR has been used for a variety of different applications such as material modeling \cite{kabliman2019prediction, kabliman2019prediction, bomarito2021development, Abdusalamov2023} or the discovery of physical relationships \cite{manzi2020orbital,gaucel2014learning, huang2021ai, sun_ouyang_zhang_zhang_2019}. \\
GP algorithms can be split up typically into four different phases: initiation, selection, evolution and termination. In the first phase an initial set of expressions is randomly created from a predefined set of possible mathematical operations, independent variables and functions. This initial set is competing in tournaments during the selection phase. In this way, random subsets are formed and the fittest individual of each subset is determined. In the next phase, the fittest individuals evolve. There are several types of mutations available e.g. \textit{crossover}, \textit{subtree}, \textit{point} or \textit{hoist mutation}. In the case of \textit{crossover} a new individual is formed from a preliminary selected parent and a donor. To this end, a random subtree of the parent is replaced by a subtree of the donor (see \autoref{fig:CrossoverExample}).
The \textit{subtree mutation} is very similar to \textit{crossover}, however, only a single parent is needed. In this case, a random subtree is replaced by a random new term allowing to reintroduce forgotten operations, functions or inputs (see for example \autoref{fig:SubtreeExample}). 
\textit{Point mutation} is an evolution of a single vertex of a tree, see \autoref{fig:PointExample}. A function, operator or input is replaced with another one. This mutation form also allows to reintroduce lost functions, operations or inputs.
The last mutation type called \textit{hoist mutation} is visualized in \autoref{fig:HoistExample}. The goal is to reduce the length of a tree. A random subtree is selected and replaced with a subtree of itself.
The selection and evolution process continues until the termination phase. A termination can happen in two ways: either a specified number of generations has been reached or a specified fitness criteria is fulfilled. Note that this procedure does not guarantee to find any optimal solutions. Nevertheless, the overall fitness of the population improves over the number of generations. Additionally, due to the random character of this approach a deterministic solution is not given. \\
For this work, the Python package \texttt{gplearn} is used \cite{stephens2019gplearn}. For most applications an additional constraint is to restrict the length of the generated expressions. Usually this is done by introducing a Lagrange multiplier as a penalty term into the calculated fitness. To discover asymptotic expansings using symbolic regression, a restriction is counterproductive for developing an asymptotic series. In \texttt{gplearn} the parsimony coefficient is responsible to keep the length of the expression small and will have a maximum value of $\SI{1e-6}{}$. For the implementation it is necessary to mention that generally symbolic regression has a high sensitivity on hyper parameters. For all examples considered in the following analytical exact solutions are used to generate artificial input data. \\
Asymptotic analysis is a powerful mathematical technique, often used to simplify complex relations for estimating the limiting behavior of interest e.g. see \cite{andrianov2002asymptotology} and references therein . This is usually done by identifying a large/small problem parameter and expanding the sought for solution in term of the series involving this parameter. In the simplest example of a given function $f(\varepsilon)$ that depends on a small parameter $\varepsilon$ the asymptotic expansion can be written as
\begin{align*}
	f(\varepsilon) &= f_0  + f_1 \varepsilon + f_2 \varepsilon^2  + f_3 \varepsilon^3 + ... \, ,
\end{align*}
where $f_i$ for $i=0,1,2, ... $ are the coefficients to be found. In the general case a series of this type can be divergent and its performance strongly depends on the value of the parameter. The evaluation of these coefficients, especially of the higher order ones is often a challenge. The goal of this paper is to adapt an SR approach for determining these coefficients as well as even the powers of the relevant parameter. Moreover, for more general expansions considered in the paper the SR approach is adapted for establishing the basics functions appearing in the asymptotic series. \\
Below, we present few examples of physically inspired problems originated from mechanics to illustrate the peculiarities of the proposed methodology. The derived SR series are compared with benchmark asymptotic expansions approximating the exact solutions of the studied problems.

\section{Collision Problem}
\label{sec:InitialImpactProblem}
In this section we will discuss an illustrative example of the collision of two bodies of mass $m_1$ and $m_2$ as shown in \autoref{fig:Balls}. For this example, mass $m_1$ has a prescribed initial velocity $v_0$, while mass $m_{2}$ is standing still. After the collision, the masses $m_1$ and $m_2$ have the velocities $v_1$  and $v_2$, respectively, which are unknown and have to be found.
The balance of linear momentum and the balance of kinetic energy are given by 
\begin{align}
	m_1 v_0 = m_1 v_1  + m_2 v_2
\end{align}
and
\begin{align}
	\frac{m_1 v_0 ^2}{2} = \frac{m_1 v_1 ^2}{2} + \frac{m_2 v_2 ^2}{2} \ .
	\label{eq:Balances2}
\end{align}
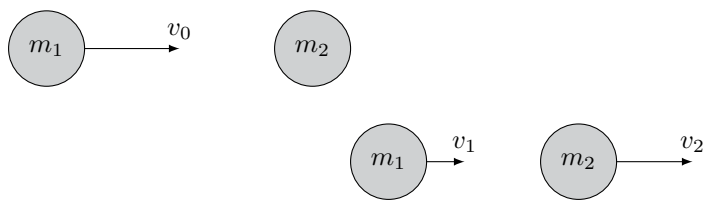
\begin{figure}[ht!]
	\centering
	\begin{tikzpicture}	
		\node[circle,draw,text=black,fill=black-25, minimum size=1cm] (c) at (0,0) {$m_1$};
		\draw [-latex](0.5,0) -- (1.75,0) node[above] {$v_0$};
		\node[circle,draw,text=black,fill=black-25, minimum size=1cm] (c) at (3.5,0) {$m_2$};
		
		\node[circle,draw,text=black,fill=black-25, minimum size=1cm] (c) at (4.5,-1.5) {$m_1$};
		\draw [-latex](5,-1.5) -- (5.5,-1.5) node[above] {$v_1$};
		\node[circle,draw,text=black,fill=black-25, minimum size=1cm] (c) at (7,-1.5) {$m_2$};
		\draw [-latex](7.5,-1.5) -- (8.5,-1.5) node[above] {$v_2$};	
	\end{tikzpicture}
	\caption{Collision problem for masses $m_1$ and $m_2$ (before and after the impact). Here $v_{0}$ is the velocity before collision while $v_{i}$ is the velocity of mass $m_{i}$ after the collission, $i=1,2$.}
	\label{fig:Balls}
\end{figure}

The solution of these equations is of the form:
\begin{align}
	u_1 &=  \frac{\delta - 1}{\delta + 1} 
	\label{eq:ExactU1}
\end{align}
and
\begin{align}
	u_2 =  \delta (1 - u_1) \ ,
\end{align}
where $\delta = \nicefrac{m_1}{m_2}$ and the dimensionless quantities $u_1 = v_1 / v_0$ and $u_2 = v_2 / v_0$. 

\begin{figure}[ht!]
	\centering
	\includegraphics[width=.48\textwidth]{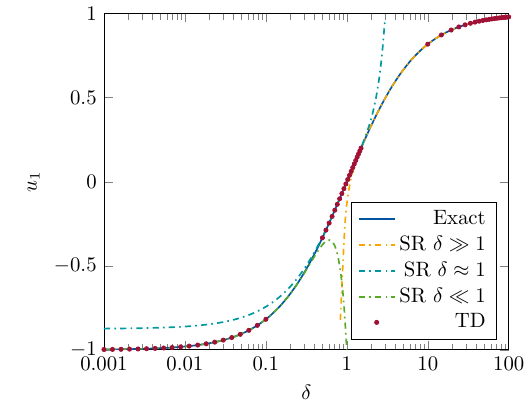}
	\caption{Training data (TD) extracted from exact solution in \autoref{eq:ExactU1} and SR asymptotic expansions determined for three limiting cases ($\delta \gg 1 $, $\delta \approx 1 $  and $\delta \ll 1$.)}
	\label{fig:Example_1_PlotApprox}
\end{figure}
\begin{figure}[ht!]
	\centering
	\includegraphics[width=.48\textwidth]{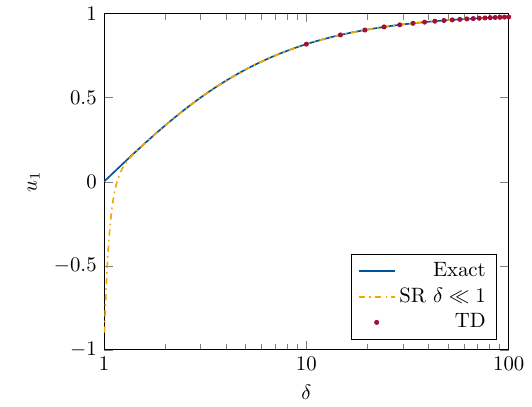}
	\caption{Training data (TD) extracted from exact solution \eqref{eq:ExactU1} and SR asymptotic series for best approximation (case $\delta \gg 1$), see \autoref{tab:P1BestApproch}.}
	\label{fig:Example_1_PlotApprox_LD}
\end{figure}

The first of these relations can be expanded into an asymptotic series for three limiting behaviors, including  $\delta \gg 1 $, $\delta \approx 1 $  and $\delta \ll 1$. The strong inequality $\delta \ll 1$ is related to the collision of a mass $m_1$ with an almost rigid wall. The case $\delta \approx 1$ corresponds to the impulse transfer through masses of almost the same weight. The strong inequality $\delta \gg 1 $ governs the collision of a large mass $m_1$ with a small mass $m_{2}$. The small parameters for each of these three scenarios are $\delta \ll 1$, $\theta = \frac{(\delta - 1)}{2} \ll 1$ and $\eta = \frac{1}{\delta} \ll 1 $. The associated converging asymptotic series become
\begin{align}
	u_{1}(\bar{\delta}) &=  -1 + 2\bar{\delta} - 2\bar{\delta}^2 + 2 \bar{\delta}^3 - 2 \bar{\delta}^4 + ... \, , \label{eq:U1Problem1LeadingOrders} \\
	u_{1}(\theta)&= \theta - \theta^2 + \theta^3 - \theta^4 + ...    \, , \label{eq:Problem1LeadingOrders}
\end{align} 
and
\begin{align}
	u_{1}(\eta) &=  1 - 2\eta + 2\eta^2 - 2\eta^3 + 2\eta^4 + ... \, .
	\label{eq:Problem1LeadingOrders2}
\end{align}
Now assume that the coefficients as well the powers of the small parameters in the series above are unknown and try to determine both of them using SR. Therefore, data is generated from the exact solution in \autoref{eq:ExactU1} for the respective domains. To this end, we implement two strategies. The first strategy starts from the chosen small parameter only, i.e. $\delta$, $\theta$ or $\eta$, specifying it as an input. Alternatively, the inputs can be given in the form of several powers of the small parameter, e.g. for series \eqref{eq:U1Problem1LeadingOrders} we can provide the input as $\{\delta, \delta^2, \delta^3\}$. Both strategies were successfully implemented, see \autoref{Appendix1}. The discussion below is mainly restricted to the first strategy due to similar outcomes for the two setups in question. Although more inputs are provided for the second strategy, the performance does not necessarily improve.\\
In \autoref{fig:Example_1_PlotApprox} we demonstrate the results for only 20 data points specified as the training data for each of limiting setups \eqref{eq:U1Problem1LeadingOrders}-\eqref{eq:Problem1LeadingOrders2}. The discovered asymptotic expansions with the best fitness are depicted. The convergence of the expansions to the exact solution is shown.
The best fits for all three series are listed in \autoref{tab:P1BestApproch}. The randomness of the initial population of the algorithm requires 5 symbolic regressions. In this case, the fitness with respect to the number of generations as well as the best fit of all 5 samples is determined  and listed in \cref{tab:P1_Fit_Small,tab:P1_Fit_Mid,tab:P1_Fit_Large}. \\
It's worth noting that all automatically constructed asymptotic expansions are close to the exact solution outside a relatively narrow domain, where the training data is given. This is clear from \autoref{fig:Example_1_PlotApprox_LD}, depicting the exact solution, along with its symbolic regression approximation and the training data. Here, only 20 data points calculated by formula \eqref{eq:ExactU1} are taken. The convergence to the exact solution for $\delta \gg 1$ and $\delta \approx 1$ is illustrated. The sought for coefficients in the determined asymptotic series given above are nearly identical to their exact values up to the \nth{17} order, see \autoref{tab:P1BestApproch}; see also the benchmark coefficients in the expansions \eqref{eq:U1Problem1LeadingOrders}-\eqref{eq:Problem1LeadingOrders2}.
\begin{table}[ht!] 
	\centering
	\caption{Best SR expansions for all three limiting cases.}
	\label{tab:P1BestApproch}
	\begin{tabular}{c p{12cm} c}
		\toprule
		Case & Best Approximation &       Fitness \\
		\midrule
		$\delta \ll 1$ & $\begin{aligned}
			u_{1}(\bar{\delta})  =&   -1.00 + 2.00\delta - 2.00\delta^2 + 2.00\delta^3 - 2.00\delta^4 + 2.00\delta^5 - 2.00\delta^6 + 2.00\delta^7 \\ &- 2.00\delta^8 + 2.00\delta^9 -2.00\delta^{10} + 2.00\delta^{11} - 2.00\delta^{12} + 2.00\delta^{13} - 2.00\delta^{14} \\ &+ 2.00\delta^{15} - 2.00\delta^{16} + 2.00\delta^{17}
		\end{aligned}$  &  \SI{5.55e-17}{}  \vspace{0.5cm} \\ 
		$\delta \approx 1$ & $\begin{aligned} u_{1}(\bar{\theta})  =&  \theta  - \theta^{2} + \theta^{3} - \theta^{4} + \theta^{5} - \theta^{6} + \theta^{7} - \theta^{8} + \theta^{9} - \theta^{10} + \theta^{11} - \theta^{12} + \theta^{13} - \theta^{14} \\ &+ \theta^{15} - \theta^{16} + \theta^{17} - \theta^{18} + \theta^{19} - \theta^{20} + \theta^{21} -\theta^{22} + \theta^{23} - \theta^{24} + \theta^{25} \end{aligned}$ & \SI{1.93e-17}{}  \vspace{0.5cm} \\
		$\delta \gg 1$ &  $\begin{aligned}  u(\eta) =&  + 1.0  - 2.0 \eta + 2.0 \eta^{2} - 2.0 \eta^{3}  + 2.0 \eta^{4} - 2.0 \eta^{5} + 2.0 \eta^{6}  - 2.0 \eta^{7} \\ &+ 2.0 \eta^{8}  - 2.0 \eta^{9}  + 2.0 \eta^{10}  - 2.0 \eta^{11} + 2.0 \eta^{12}  - 2.0 \eta^{13}  + 2.0 \eta^{14} \\ &- 2.0 \eta^{15} + 2.0 \eta^{16} - 2.0 \eta^{17}   \end{aligned}$ & \SI{3.70e-17}{} \\
		\bottomrule  
	\end{tabular}
\end{table}

\section{Kelvin-Voigt Viscoelastic Solid}
\label{sec:KelvinVoigtModel}
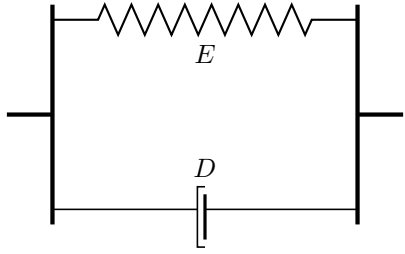
\begin{figure}[ht!]
	\centering
	\begin{tikzpicture}[
		node distance = 0mm,
		start chain = going right,
		box/.style = {draw,
			font=\linespread{0.75}\selectfont\small,
			align=center, inner sep=2mm, outer sep=0pt,
			on chain},
		axs/.style = {draw, minimum width=12mm, minimum height=2mm,
			inner sep=0pt, outer sep=0pt,
			on chain, node contents={}},
		arr/.style = {color=#1, line width=0.8mm,
			shorten >=-1mm, shorten <=-1mm,
			-{Stealth[length=1.6mm,width=3mm,flex=1.2]},
			bend angle=60},
		spring/.style = {thick, decorate,       
			decoration={zigzag,amplitude = 2mm, pre length=6mm,post length=6mm,segment length=10}
		},
		damper/.pic = {\coordinate (-east);   
			\coordinate[left=1mm of -east] (-west);
			\draw[black, very thick] ($(-east)+(0,2mm)$) -- ++ (0,-6mm);
			\draw[black, semithick]  ($(-east)+(0,3mm)$) -| ++ (-1mm,-8mm) -- ++ (1mm,0);
		},
		shorten <>/.style = {shorten >=#1, shorten <=#1},
		]
		
		\node (n5) [black, minimum height=25mm,minimum width=40mm,
		label={[yshift= -7mm,black]above:$E$},
		label={[black, yshift=8mm]below:$D$},
		on chain] {};
		\draw[black, ultra thick,shorten <>=-2mm] (n5.north west) -- (n5.south west);
		\draw[black, ultra thick,shorten <>=-2mm] (n5.north east) -- (n5.south east);
		\draw[black, spring]   (n5.north west) -- (n5.north east);
		\pic (dmp) at (n5.south)  {damper};
		\draw[black, semithick]  (n5.south west) -- (dmp-west)  (dmp-east) -- (n5.south east);
		
		\draw[black, ultra thick,shorten <>=-2mm] (n5.west) ++ (-2mm,0mm) -- ++ (-2mm,0mm);
		\draw[black, ultra thick,shorten <>=-2mm] (n5.east) ++ (2mm,0mm) -- ++ (2mm,0mm);
		
	\end{tikzpicture}
	\caption{Kelvin-Voigt solid with a spring of stiffness $E$ and a damper of viscosity $D$.}
	\label{fig:KelvinVoigtModel}
\end{figure}

Next consinder a more evolved example arising from the viscoelastic Kelvin-Voigt model, e.g. see \cite{christensen2012theory}. Let us start from the constitutive relation (\autoref{fig:KelvinVoigtModel}) 
\begin{align}
	\sigma = E \varepsilon + D \frac{d \varepsilon}{d t} \, . \label{eq:StrainPDEVoigt}
\end{align}
where  $\sigma$ is the stress, $\varepsilon$ is the strain, $E$ is the stiffness of a spring (Young's modulus) and $D$ the viscosity of a damper while $t$ denotes time. This formula can be rewritten in an integral form as
\begin{align*}
	\varepsilon(t) &= \varepsilon(0) \exp(- \theta t) + \frac{\exp(-\theta t)}{D}  \int_{t_{1} = 0}^{t} \sigma(t_{1}) \exp(\theta t_{1}) \mathop{d t_{1}} \, ,
\end{align*}
with $\theta = \nicefrac{E}{D}$. Introduce a typical time scale $T$ assuming that $t \gg T$. The limiting large time behavior of the last formula becomes
\begin{align}
	\varepsilon(\tau) &= \varepsilon(0) \exp(- \delta \tau) + \frac{T}{D}  \exp(-\delta \tau) \int_{\tau_{1} = 0}^{\infty} \sigma(\tau_{1}) \exp(\delta \tau_{1}) \mathop{d \tau_{1}} \ ,
	\label{eq:EpsInt}
\end{align}
where $\delta = \theta T$ and $\tau = \nicefrac{t}{T}$.

\begin{figure}[ht!]
	\centering
	\begin{subfigure}[b]{0.49\textwidth}
		\centering
		\includegraphics[width=\textwidth]{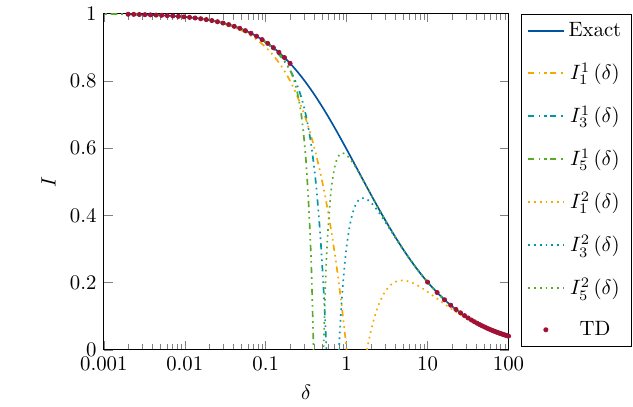}
		\caption{Benchmark expansions}
		\label{fig:Example_2_PlotApproxBM}
	\end{subfigure}
	\hfill
	\begin{subfigure}[b]{0.49\textwidth}
		\centering
		\includegraphics[width=\textwidth]{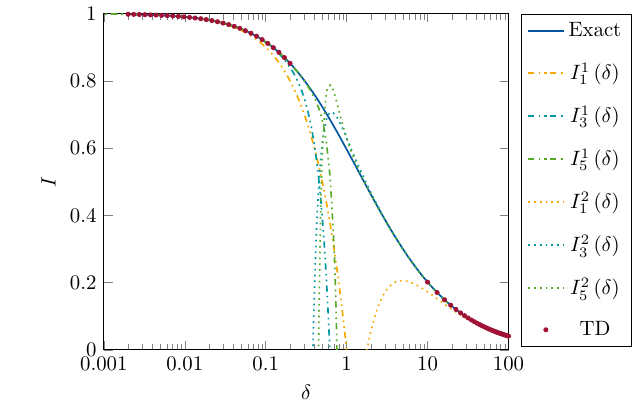}
		\caption{SR expansions}
		\label{fig:Example_2_PlotApproxSR}
	\end{subfigure}
	\caption{ (a) Benchmark and (b) SR asymptotic expansions $I_{i}^{1}(\delta)$ and $I_{i}^{2}(\deltat)$  for $\deltat \ll 1$ and  $\deltat \gg 1$, respectively; index $i$ denotes the highest order in Equations \ref{eq:P2_Small} and  \ref{eq:P2_Large}.}
	\label{fig:Example_2_PlotApprox}
\end{figure}

Next assume that the studied Kelvin-Voigt solid is loaded by the stress depending on time $\tau$ as follows
\begin{align*}
	\tilde{\sigma}(\tilde{\tau}) =   \frac{\sigma_{0}\exp((1 - \tilde{\delta}) \tilde{\tau} )}{1 + \tilde{\delta} \tilde{\tau}} \, ,
\end{align*}
where $\sigma_{0}$ is a prescribed amplitude. In this case, formula \eqref{eq:EpsInt} can be reduced to
\begin{align*}
	\varepsilon(t) &= \left(\varepsilon(0) + \frac{T\sigma_{0} }{D} I(\delta)\right) \exp(- \delta \tau)  \ ,
\end{align*}
with (see \cite{prudnikov1986integrals})
\begin{align}
	I(\delta) =	\int_{0}^{\infty} \frac{\exp (x) }{1+ \tilde{\delta} x } \mathop{d x} = \frac{ e^{\nicefrac{1}{\delta}} }{\delta} \Gamma \left(0, \frac{1}{\delta}\right) \ ,
	\label{eq:P2Exact}
\end{align}
where $\Gamma$ denotes the Gamma function. \\
The integral $I(\delta)$ can be expanded into two different asymptotic series at  $\delta \ll 1$and $\delta \gg 1$. However, in contrast to the previous example the first of them appears to be divergent. The small parameters for each of these scenarios are $\delta \ll 1$ and $\eta = \frac{1}{\delta} \ll 1 $. They are given by, e.g. see \cite{erdelyi1956asymptotic} and references therein,
\begin{align}
	I(\delta) &= 1 - \delta + 2 \delta^2 - 6 \delta^3 + 24 \delta^4 + ... \ , \label{eq:P2_Small}
\end{align}
and 	
\begin{align}
	I(\eta) =& \eta (-\gamma - \log{\eta})  + \eta^2 (1 - \gamma - \log{\eta})  + \frac{\eta^3}{4} (3 - 2 \gamma - 2 \log{\eta}) + \frac{\eta^4}{36} (11 - 6 \gamma - 6 \log{\eta}) + ...  \, . \label{eq:P2_Large}
\end{align}
Therein, $\gamma$ is the Euler–Mascheroni constant. In this section we adapt the symbolic regression for the construction of the analogs of the series in \eqref{eq:P2_Small} and \eqref{eq:P2_Large} using the exact formula for the integral $I(\delta)$ for generating training data. The main focus below is on the effect of the divergent behavior of \autoref{eq:P2_Small}. The numerical results are displayed in \autoref{fig:Example_2_PlotApprox}. Note that for the SR expansions for the case $\eta = \frac{1}{\delta} \ll 1 $ the inputs have been provided with $\{\eta, \log \eta\}$.

As expected, in contrast to convergent series, there is a natural threshold for the number of terms in divergent series which can be reproduced with a required accuracy  over the chosen domain of the small parameter, see \autoref{tab:P2BestApproch}. The SR higher order terms strongly deviate from their counterparts in  benchmarked asymptotic divergent expansions, see also Equations \ref{eq:P2_Small} and \ref{eq:P2_Large}. This is quite obvious, as symbolic regression aims at achieving the best possible accuracy of the provided data, which is not always feasible when using divergent asymptotic series.

\begin{table}[ht!] 
	\centering
	\caption{Best SR expansions for all three limiting cases.}
	\label{tab:P2BestApproch}
	\begin{tabular}{c p{12cm} c}
		\toprule
		Case & Best Approximation &       Fitness \\
		\midrule
		$\delta \ll 1$ & $\begin{aligned} I(\delta) =&  1.0- 1.0 \delta + 1.95 \delta^{2} - 4.74\delta^{3} + 9.16\delta^{4} - 9.12\delta^{5} - 1.48\delta^{6} + 19.14\delta^{7} \\ &- 15.34\delta^{8}  \end{aligned}$ & \SI{6.01e-06}{}  \vspace{0.5cm} \\
		$\delta \gg 1$ &  $\begin{aligned}  I(\eta) =& \eta (- \log{\eta} - 0.58) + \eta^{2}  (0.42 - 1.0 \log{\eta}) + \eta^{3}  (0.01 \log{\eta}^{2} - 0.38 \log{\eta} + 0.79)  \\ &+ \eta^{4} (- 0.06 \log{\eta}^{2} + 0.09 \log{\eta}) + \eta^{5}  (0.01 \log{\eta}^{3} - 0.03 \log{\eta}^{2} - 0.01 \log{\eta}) \\ &+ \eta^{6} (- 0.01 \log{\eta}^{4} - 0.04 \log{\eta}^{3} + 0.02 \log{\eta}^{2}) + 0.02 \eta^{7} \log{\eta}^{2}     \end{aligned}$ & \SI{2.76e-09}{} \\
		\bottomrule  
	\end{tabular}
\end{table}

\begin{figure}[ht!]
	\centering
	\begin{subfigure}{0.49\textwidth}
		\includegraphics[trim=100 65 20 45, clip, width=0.95\textwidth]{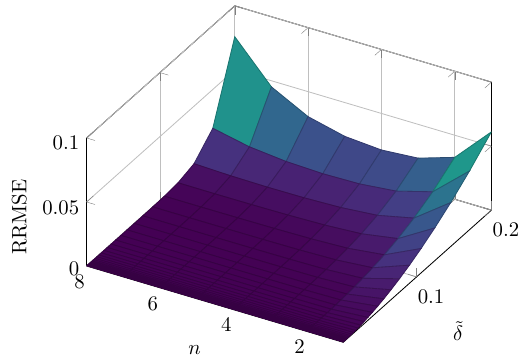}
		\caption{}
		\label{fig:Example_2_ErrorAsympAna}
	\end{subfigure}
	\hfill
	\begin{subfigure}{0.49\textwidth}
		\includegraphics[trim=100 65 20 45, clip, width=0.95\textwidth]{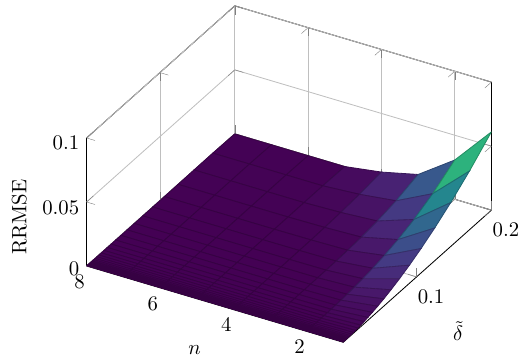}
		\caption{}
		\label{}
	\end{subfigure}
	\caption{The relative root mean square error  (RRMSE) of benchmark and SR expansions, see \eqref{eq:P2_Small} and  \eqref{eq:P2_Large} as well as \autoref{tab:P2BestApproch}, plotted against exact solution for different orders $n$ and values of $\delta$ .}
	\label{fig:Example2Erros}
\end{figure}

\begin{figure}[ht!]
	\centering
	\includegraphics[width=0.6\textwidth]{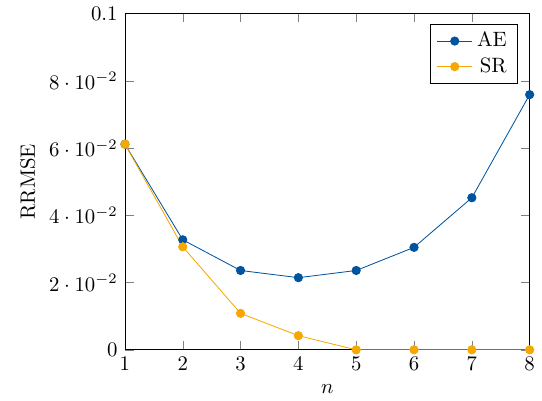}
	\caption{The relative root mean square error  (RRMSE) of benchmark analytic expansion (AE) and SR for $\deltat = \SI{0.2}{}$ plotted against exact solution \eqref{eq:P2Exact}. }
	\label{fig:Example_2_ErrorCompare}
\end{figure}

The deviation of the benchmark and SR expansions from the exact solution is plotted using the relative root mean square error (RRMSE) in \autoref{fig:Example2Erros} vs. the parameter $\delta$ and the highest order $n$ of the retained term in the analyzed series for the case $\delta \ll 1$.
In \autoref{fig:Example2Erros} and  also in the next \autoref{fig:Example_2_ErrorCompare}, the range of the small problem parameter $\deltat$ is specified as $\SI{2e-4}{} \le \delta \le \SI{0.2}{}$ while $n \le 8$.  It appears that there exists an optimal order $n$ for the analytical divergent series related to the most accurate approximation at the given value of $\delta$. In particular, $n=4$ in \autoref{fig:Example_2_ErrorCompare} corresponds to $\delta = 0.2$.
In this case, however, the accuracy of the SR series is higher approaching a plateau as the order $n$ increases.

\section{Elastic Bending Wave}
\label{sec:ElasticLayer}
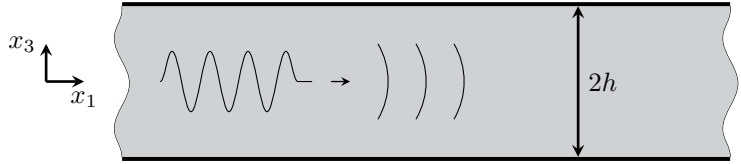
\begin{figure}[ht!]
	\centering
	\begin{tikzpicture}
		\draw[line width = 3, name path = A]  (0,0) -- (8, 0);
		\draw[line width = 3, name path = B]  (0,2) -- (8, 2);
		\path [draw, decorate, decoration={snake, segment length=12mm, amplitude=1mm}, name path = C] (0,-0.) -- (0,2.);
		\path [draw, decorate, decoration={snake, segment length=12mm, amplitude=1mm}, name path = D] (8,-0.) -- (8,2.);
		\tikzfillbetween[of=C and D]{black-25};
		\draw[line width = 1,-stealth]  (-1,1) -- (-1, 1.5) node[left] {$x_3$};
		\draw[line width = 1,-stealth]  (-1,1) -- (-0.5, 1.) node[below] {$x_1$};
		\draw[line width = 1,stealth-stealth]  (6,0) -- node[right] {$2h$} (6, 2);
		\path [draw, decorate, decoration={snake, segment length=5mm, amplitude=4mm}, black] (0.5,1) -- (2.5,1.);
		\draw[line width=0.5, -stealth, black] (2.75,1) -- (3.,1.);
		\draw[black] ([shift=(-30:1cm)]2.5,1) arc (-30:30:1cm);
		\draw[black] ([shift=(-30:1cm)]3,1) arc (-30:30:1cm);
		\draw[black] ([shift=(-30:1cm)]3.5,1) arc (-30:30:1cm);
	\end{tikzpicture} 
	\caption{Rayleigh-Lamb waves schematic in an elastic layer.} 
	\label{fig:RLPlate}
\end{figure}

As the final example, we consider Rayleigh-Lamb waves propagating along an elastic layer of thickness $2h$ with traction free faces (\autoref{fig:RLPlate}), see the original papers \cite{lamb1917waves, rayleigh1888free} . The equation of motion in cartesian coordinates $x_{1},x_{2},x_{3}$ is given by (here and below in this section see \cite{kaplunov1998dynamics} for more details)
\begin{align*}
	\frac{E}{2 (1+ \nu)} \Delta \tens{u} + \frac{E}{2(1 + \nu)(1 - 2\nu)} \nabla(\nabla \cdot \tens{u}) - \rho \frac{\partial^2 \tens{u}}{\partial t^2} = 0 \, ,
\end{align*}
where $E$ is Young's modulus, $\nu$ is Poisson's ratio, $\rho$ the mass density and $t$ time. We restrict ourselves to the plane-strain problem in the plane $x_{1}$-$x_{3}$. In this case, the displacement vector is given by $\tens{u} = (u_{1}, 0, u_{3})$. Then the boundary conditions along the faces $x_{3} = \pm h$ become
\begin{align*}
	\frac{\partial u_{3}}{\partial x_{1}} + \frac{\partial u_{1}}{\partial x_{3}} &=  \frac{\nu}{1-\nu}\frac{\partial u_{1}}{\partial x_{1}} + \frac{\partial u_{3}}{\partial x_{3}}= 0 \, .
\end{align*} 


The associated dispersion relation for antisymmetric traveling waves with angular frequency $\omega$ and wavenumber $k$ can be written as
\begin{align}
	\gamma^4 \frac{\sinh \alpha }{\alpha} \cosh \beta  - \beta^2 K^2 \cosh \alpha \frac{\sinh \beta}{\beta} = 0 \, , \label{eq:P3Exact}
\end{align}
with 
\begin{align*}
	\gamma^2 &= K^2 - \frac{1}{2} \Omega^2 , &  \alpha^2 &= K^2 - \kappa^2 \Omega^2 & &\text{and} & \beta^2 &= K^2 - \Omega^{2} \, ,
\end{align*}
where the dimensionless circular frequency $\Omega = \nicefrac{\omega h}{c_{2}}$, the dimensionless wavenumber $K = kh$, the shear wave speed $c_{2} = \sqrt{\frac{E}{2(1+\nu)\rho}}$ and $\kappa = \sqrt{\frac{1-2\nu}{2-2\nu}}$. 
This equation can be solved numerically, e.g. using a nonlinear solver in Python. Below the numerical results generated by Python, are used as training data. We also present the low-wave frequency of the fundamental antisymmetric, i.e. bending, mode see also \cite{goldenveizer1993timoshenko, gol1990asymptotic}, given by
\begin{align}
	K^{4}_{n} = \frac{3}{2} (1 - \nu) \Omega^{2} \sum_{j=0}^{n} A_{j} \Omega^{j} \, , \label{eq:Example3Approx}
\end{align}
where the first four coefficients $A_{j}$ take the form 
\begin{align*}
	A_{0} &= 1 , & A_{1} &= \chi \frac{17 -7\nu}{15(1-\nu)} , \\
	A_{2} &= \frac{1179 -818\nu +409\nu^{2}}{2100(1 - \nu)}, & A_{3} &= \chi \frac{5951 - 2603\nu +9953\nu^2 - 4901\nu^3}{126000(1-\nu)^2} \, ,
\end{align*}
with $\chi = \sqrt{\nicefrac{3(1-\nu)}{2}}$. 

%

\begin{figure}[ht!]
	\centering
	\includegraphics[width=.5\textwidth]{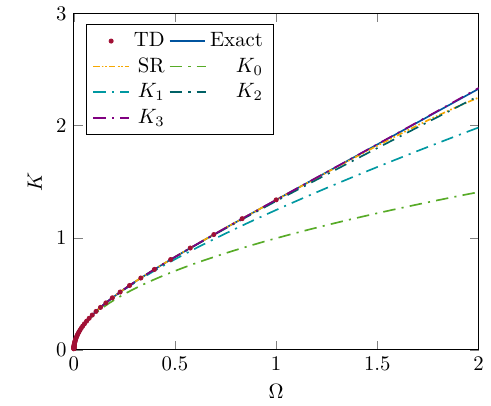}
	\caption{Exact solution of \autoref{eq:P3Exact} along with its asymptotic expansions 
		$K_{n}^4$ by \autoref{eq:Example3Approx} 
		and SR expansion.}
	\label{fig:Example_3_Plot}
\end{figure}

The best SR appproximation is depicted in \autoref{fig:Example_3_Plot} along with the exact solution and the asymptotic series from formula \eqref{eq:Example3Approx} at the orders $n=0,1,2,3$. The aforementioned SR approximation has the same coefficients at $n=3$. Two latter are computed at $\nu = \SI{0.3455}{}$, whereas Poisson's ratio is not specified as an input for the SR approximation. Note that using the determined SR values of the coefficient $A_{1}$ or $A_{2}$  we may restore the unknown Poisson ratio by the following formulae
\begin{align}
	\nu(A_{1}) = \frac{119-75 A_{1}^2 + 5 \sqrt{15} \sqrt{15 A_{1}^4 - 28 A_{1}^2} }{49} 
\end{align}
or
\begin{align}
	\nu(A_{2}) = \frac{409 - 1050 A_{2} + \sqrt{70} \sqrt{15750 A_{2}^2 - 4499} }{409} .
\end{align}
This seems promising for the evaluation of Poisson's ratio from experimental data, e.g. see \cite{kaplunov1992determination, rogers1995elastic}. The convincing results for the considered example are exposed in \autoref{tab:P3Poissons}. 

\begin{table}[ht!] 
	\centering
	\caption{Determination of Poisson's ratio from coefficients of SR expansions.}
	\label{tab:P3Poissons}
	\begin{tabular}{c c c c c c c c}
		\toprule
		& 1 & 2 & 3 & 4 & 5 & Mean & $\nu(A_i)$ \\
		\midrule
		$A_{1}$ & 1.48 & 1.56  & 1.35 & 1.59 & 1.42 & 1.48 & 0.36 \\
		$A_{2}$ & 0.60 & 0.57 & 0.97 & 0.54 & 0.81 & 0.71 & 0.38 \\
		\bottomrule  
	\end{tabular}
\end{table}

\section{Concluding Remarks}
\label{sec:Conclusions}
In this paper we developed a robust framework to obtain SR asymptotic expansions, illustrated by three problems in mechanics. The proposed methodology was first adapted for an initial simple setup of a two-mass collision problem resulting in the asymptotic expansions expressed through the convergent series corresponding to three limiting behaviors. The SR approximations matched the benchmark analytic expansions up to the \nth{17} order for all mass ratios despite the well-known sensitivity of SR hyperparameters. The example of a Kelvin-Voigt solid is considered to illustrate the peculiarities of SR series for an originally divergent asymptotic expansion. Its observed that the accuracy of an SR expansion makes it superior over the associated optimal analytic series. The last example is concerned with a bending wave propagating along an elastic layer. The long-wave low frequency asymptotic behavior was tackled. A possibility of using the obtained SR results for the evaluation of the unknown Poisson ratio is indicated. \\
An "asymptotic" way of thinking is adapted for SR implementation. Although all the training data in this paper are generated using explicit exact solutions, this approach can equivalently be applied to data captured from experiments. Another natural possibility is to make use of numerical data, e.g. FEM simulations. It is remarkable that the obtained SR expansions were discovered from very few data points in comparison with alternative ML techniques. Above we restricted ourselves to basic types of asymptotic behavior. A follow up program may include multi-parametric asymptotic analysis, matched asymptotic expansions as well as Pad\'{e} approximations and other more involved techniques.


\section*{Acknowledgement}
Julius Kaplunov gratefully acknowledges the support of the Alexander von Humboldt Foundation which made possible his three months visit to the Department of Continuum Mechanics at RWTH Aachen University in summer 2021.

\newpage
\section*{Appendix}
\section*{Collision Problem}
\label{Appendix1}
\begin{table}[H]
	\centering
	\caption{SR expansions for 5 samples ($\delta \ll 1$).}
	\label{tab:P1_Fit_Small}
	\begin{tabular}{c p{13cm} c}
		\toprule
		N & Best Approximation &       Fitness \\
		\midrule
		1 & $\begin{aligned} u_{1}(\bar{\delta})  =& -1.00 + 2.00\delta - 2.00\delta^2 + 2.00\delta^3 - 2.00\delta^4 + 2.00\delta^5 - 2.00\delta^6 + 2.00\delta^7 - 2.00\delta^8 \\ &+ 2.00\delta^9 
			-2.00\delta^{10} + 2.00\delta^{11} - 2.27\delta^{12} + 4.55\delta^{13} - 4.55\delta^{14} + 4.55\delta^{15} + 9.05\delta^{16} \\ &- 9.05\delta^{17} + 36.04\delta^{19}
		\end{aligned}$  &  \SI{2.597e-15}{} \vspace{0.5cm} \\
		2 & $\begin{aligned} u_{1}(\bar{\delta})  =& -1.00 + 2.00\delta -2.00\delta^2 + 2.00\delta^3 - 2.00\delta^4 + 2.00\delta^5 - 2.00\delta^6 + 2.00\delta^7 - 2.00\delta^8 \\ &+ 2.00\delta^9 - 2.00\delta^{10} + 2.00\delta^{11}  \end{aligned}$  &  \SI{5.977e-15 }{} \vspace{0.5cm}\\
		3 &$\begin{aligned} u_{1}(\bar{\delta})  =&  - 1.00 + 2.00 \delta - 2.01 \delta^{2} +1.97 \delta^{3}  \end{aligned}$  & \SI{5.128e-06 }{} \vspace{0.5cm} \\
		4 & $\begin{aligned}
			u_{1}(\bar{\delta})  =&   -1.00 + 2.00\delta - 2.00\delta^2 + 2.00\delta^3 - 2.00\delta^4 + 2.00\delta^5 - 2.00\delta^6 + 2.00\delta^7 - 2.00\delta^8 \\ &+ 2.00\delta^9 -2.00\delta^{10} + 2.00\delta^{11} - 2.00\delta^{12} + 2.00\delta^{13} - 2.00\delta^{14} + 2.00\delta^{15} - 2.00\delta^{16} \\ &+ 2.00\delta^{17}
		\end{aligned}$  &  \SI{5.551e-17}{}  \vspace{0.5cm} \\
		5 & $\begin{aligned}
			u_{1}(\bar{\delta})  =&   -1.00 + 2.00\delta - 1.99\delta^2 + 1.7137145\delta^3  + 0.01\delta^5 - 0.01\delta^9 
		\end{aligned}$  &  \SI{8.108e-07}{}  \\
		\bottomrule  
	\end{tabular}
\end{table}

\begin{table}[H]
	\centering
	\caption{SR expansions for 5 samples ($\delta \approx 1$).}
	\label{tab:P1_Fit_Mid}
	\begin{tabular}{c p{13cm} c}
		\toprule
		N & Best Approximation &       Fitness \\
		\midrule
		1 & $\begin{aligned} u_{1}(\bar{\theta})  =&  \theta  - \theta^{2} + \theta^{3} - \theta^{4} + \theta^{5} - \theta^{6}  + \theta^{7} - \theta^{8} + \theta^{9} - \theta^{10} + \theta^{11} - \theta^{12} + \theta^{13} - \theta^{14} + \theta^{15} \\ &- \theta^{16} + \theta^{17} - \theta^{18} + \theta^{19} - \theta^{20} + \theta^{21}  - \theta^{22} \end{aligned}$ & \SI{1.82e-15}{}  \vspace{0.5cm} \\
		2 & $\begin{aligned} u_{1}(\bar{\theta})  =&  \theta  - \theta^{2} + \theta^{3} - \theta^{4} + \theta^{5} - \theta^{6}  + \theta^{7} - \theta^{8} + \theta^{9} - \theta^{10} + \theta^{11} - \theta^{12} + \theta^{13} - \theta^{14} + \theta^{15} \\ &- \theta^{16} + \theta^{17} - \theta^{18} + \theta^{19} - \theta^{20} + \theta^{21}  - \theta^{22} +\theta^{23} \end{aligned}$ & \SI{1.18e-16}{}  \vspace{0.5cm} \\
		3 & $\begin{aligned}  u_{1}(\bar{\theta})  =& \theta  - \theta^{2} + \theta^{3} - \theta^{4}  + \theta^{5} - \theta^{6} + \theta^{7} - \theta^{8} + \theta^{9}  - \theta^{10} + \theta^{11} - \theta^{12} + \theta^{13} - \theta^{14} + \theta^{15} \\ &- \theta^{16} + \theta^{17} - \theta^{18} + \theta^{19} - \theta^{20} + \theta^{21} - \theta^{22} + \theta^{23}   \end{aligned}$ & \SI{4.52e-16}{}  \vspace{0.5cm} \\
		4 & $\begin{aligned} u_{1}(\bar{\theta})  =&  \theta  - \theta^{2} + \theta^{3} - \theta^{4} + \theta^{5} - \theta^{6} + \theta^{7} - \theta^{8} + \theta^{9} - \theta^{10} + \theta^{11} - \theta^{12} + \theta^{13} - \theta^{14} + \theta^{15} \\ &- \theta^{16} + \theta^{17} - \theta^{18} + \theta^{19} - \theta^{20} + \theta^{21} \end{aligned}$ & \SI{7.35e-15}{}  \vspace{0.5cm} \\
		5 & $\begin{aligned} u_{1}(\bar{\theta})  =&  \theta  - \theta^{2} + \theta^{3} - \theta^{4} + \theta^{5} - \theta^{6} + \theta^{7} - \theta^{8} + \theta^{9} - \theta^{10} + \theta^{11} - \theta^{12} + \theta^{13} - \theta^{14} + \theta^{15} \\ &- \theta^{16} + \theta^{17} - \theta^{18} + \theta^{19} - \theta^{20} + \theta^{21} -\theta^{22} + \theta^{23} - \theta^{24} + \theta^{25} \end{aligned}$ & \SI{1.93e-17}{}  \\
		\bottomrule  
	\end{tabular}
\end{table}
\begin{table}[H]
	\centering
	\caption{SR expansions for 5 samples ($\delta \gg 1$).}
	\label{tab:P1_Fit_Large}
	\begin{tabular}{c p{13cm} c}
		\toprule
		N & Best Approximation &       Fitness \\
		\midrule
		1 &  $\begin{aligned}  u_{1}(\eta) =& 1.0 - 2.0 \eta+ 2.0 \eta^{2}- 2.0 \eta^{3}+ 2.0 \eta^{4} - 2.0 \eta^{5} + 2.0 \eta^{6} - 2.0 \eta^{7} + 2.0 \eta^{8} - 2.0 \eta^{9} \\ &+ 2.0 \eta^{10} - 2.0 \eta^{11}  + 2.0 \eta^{12}   - 2.0 \eta^{13}      \end{aligned}$ & \SI{1.23e-17}{}  \vspace{0.5cm} \\
		2 &  $\begin{aligned} u_{1}(\eta) =&  + 1.0  - 2.0 \eta + 2.0 \eta^{2} - 2.0 \eta^{3} + 2.0 \eta^{4} - 2.0 \eta^{5} + 2.0 \eta^{6} - 2.0 \eta^{7}  + 2.0 \eta^{8}- 2.0 \eta^{9}    \\ &+ 2.0 \eta^{10}  - 2.0 \eta^{11}   + 2.0 \eta^{12} - \eta^{13}  \end{aligned}$ & \SI{4.68e-15}{}  \vspace{0.5cm} \\
		3 &  $\begin{aligned}  u_{1}(\eta) =& + 1.0 - 2.0 \eta   + 2.0 \eta^{2} - 2.0 \eta^{3} + 2.0 \eta^{4} - 2.0 \eta^{5} + 2.0 \eta^{6} - 2.0 \eta^{7} + 2.0 \eta^{8} - 2.0 \eta^{9}  \\ &+ 2.0 \eta^{10} - 2.0 \eta^{11}    \end{aligned}$ & \SI{6.78e-17}{}  \vspace{0.5cm} \\
		4 &  $\begin{aligned}  u_{1}(\eta) =&  + 1.0  - 2.0 \eta + 2.0 \eta^{2} - 2.0 \eta^{3}  + 2.0 \eta^{4} - 2.0 \eta^{5} + 2.0 \eta^{6}   - 2.0 \eta^{7} + 2.0 \eta^{8}  - 2.0 \eta^{9}  \\ &+ 2.0 \eta^{10}  - 2.0 \eta^{11} + 2.0 \eta^{12}  - 2.0 \eta^{13}  + 2.0 \eta^{14} - 2.0 \eta^{15} + 2.0 \eta^{16} - 2.0 \eta^{17}   \end{aligned}$ & \SI{3.70e-17}{}  \vspace{0.5cm} \\
		5 &  $\begin{aligned}    u_{1}(\eta) =&  + 1.0 - 2.0 \eta + 2.0 \eta^{2} - 2.0 \eta^{3} + 2.0 \eta^{4} - 2.0 \eta^{5}   + 2.0 \eta^{6} - 2.0 \eta^{7}+ 2.0 \eta^{8} - 2.0 \eta^{9} \\ &+ 2.0 \eta^{10} - 2.0 \eta^{11} + 2.0 \eta^{12} - 2.0 \eta^{13}   \end{aligned}$ & \SI{1.02e-15}{}  \\
		\bottomrule  
	\end{tabular}
\end{table}
\begin{table}[H]
	\centering
	\caption{SR expansions for 5 samples ($\delta \ll 1$) for inputs $\{\delta, \delta^2, \delta^3\}$.}
	\label{tab:P1_Fit_Small_HO}
	\begin{tabular}{c p{13cm} c}
		\toprule
		N & Best Approximation &       Fitness \\
		\midrule
		1 & $\begin{aligned} u_{1}(\bar{\delta})  =&  - 1.0 + 2.0 \delta - 2.0 \delta^{2}+ 2.0 \delta^{3} - 2.0 \delta^{4} + 2.0 \delta^{5}  - 2.0 \delta^{6} + 2.0 \delta^{7} - 2.0 \delta^{8} + 2.0 \delta^{9} \\ &- 2.0 \delta^{10} + 2.0 \delta^{11} - 1.0 \delta^{12} + \delta^{13}  \end{aligned}$  &  \SI{3.11e-15}{} \vspace{0.5cm} \\
		2 & $\begin{aligned} u_{1}(\bar{\delta})  =& - 1.0  + 2.0 \delta  - 2.0 \delta^{2} + 2.0 \delta^{3} - 2.0 \delta^{4}  + 2.0 \delta^{5} - 2.0 \delta^{6} + 2.0 \delta^{7}  - 2.0 \delta^{8} + 2.0 \delta^{9} \\ &- 2.0 \delta^{10} + 2.0 \delta^{11} - 2.0 \delta^{12}  + 2.0 \delta^{13}  - 2.0 \delta^{14}   \end{aligned}$  &  \SI{1.48e-16 }{} \vspace{0.5cm} \\
		3 & $\begin{aligned} u_{1}(\bar{\delta})  =& - 1.0  + 2.0 \delta - 2.0 \delta^{2} + 2.0 \delta^{3} - 2.0 \delta^{4} + 2.0 \delta^{5} - 2.0 \delta^{6} + 2.0 \delta^{7} - 2.0 \delta^{8} + 2.0 \delta^{9}  \\ &- 2.0 \delta^{10}   + 2.0 \delta^{11}  - 2 \delta^{12}  + 3.0 \delta^{13}  - 2.0 \delta^{14}  + \delta^{15}   - \delta^{16} \end{aligned}$  &  \SI{5.79e-15}{} \vspace{0.5cm} \\
		4 & $\begin{aligned} u_{1}(\bar{\delta})  =&   - 1.0  + 2.0 \delta  - 2.0 \delta^{2} + 2.0 \delta^{3}  - 2.0 \delta^{4}+ 2 \delta^{5} - 2.0 \delta^{6}   + 2.0 \delta^{7}   - 2.0 \delta^{8}   + 2.0 \delta^{9} \\ &- \delta^{10} \end{aligned}$  &  \SI{4.25e-14}{} \vspace{0.5cm} \\
		5 & $\begin{aligned} u_{1}(\bar{\delta})  =&  - 1.0 + 2.0 \delta  - 2 \delta^{2} + 2.0 \delta^{3} - 2.0 \delta^{4}  + 2.0 \delta^{5} - 2.0 \delta^{6} + 2.0 \delta^{7} - 2.0 \delta^{8} + 2.0 \delta^{9}   \\ &- 2.28 \delta^{10}  + 8.85 \delta^{11}  - 11.85 \delta^{12}   + 5.28 \delta^{13}  + 2.0 \delta^{14}  - 3.0 \delta^{15}  +\delta^{16}  \end{aligned}$  &  \SI{4.69e-15}{} \\
		\bottomrule  
	\end{tabular}
\end{table}
\begin{table}[H]
	\centering
	\caption{SR expansions for 5 samples ($\delta \approx 1$) for inputs $\{\delta, \delta^2, \delta^3\}$.}
	\label{tab:P1_Fit_Mid_HO}
	\begin{tabular}{c p{13cm} c}
		\toprule
		N & Best Approximation &       Fitness \\
		\midrule
		1 & $\begin{aligned} u_{1}(\bar{\theta})  =& 1.0 \theta - 1.0 \theta^{2} + 1.0 \theta^{3} - 1.0 \theta^{4} + 1.0 \theta^{5}  - 1.0 \theta^{6}  + 1.0 \theta^{7} - 1.0 \theta^{8} + 1.0 \theta^{9} \\ &- 1.0 \theta^{10} + 1.0 \theta^{11}  - 1.0 \theta^{12} + 1.0 \theta^{13} - 1.0 \theta^{14} + 1.0 \theta^{15} - 1.0 \theta^{16} + 1.0 \theta^{17} \\ &- 1.0 \theta^{18} + 1.0 \theta^{19} - 1.0 \theta^{20} + 1.0 \theta^{21} \end{aligned}$ & \SI{7.344916e-15}{}  \vspace{0.5cm} \\
		2 & $\begin{aligned}   u_{1}(\bar{\theta})  =&  \theta - \theta^{2} + \theta^{3} - \theta^{4} + \theta^{5} - \theta^{6} + \theta^{7} - \theta^{8}  + \theta^{9}   - \theta^{10} + \theta^{11} - \theta^{12} + \theta^{13} - \theta^{14}  \\ &+ \theta^{15} - \theta^{16} + \theta^{17} \- \theta^{18}+ \theta^{19} - \theta^{20} + \theta^{21}   - \theta^{22}   \end{aligned}$ & \SI{7.699281e-16}{}  \vspace{0.5cm} \\
		3 & $\begin{aligned}  u_{1}(\bar{\theta})  =&  \theta  - \theta^{2} + \theta^{3}- \theta^{4} + \theta^{5}  - \theta^{6} + \theta^{7} - \theta^{8} + \theta^{9}  - \theta^{10} + \theta^{11} - \theta^{12} + \theta^{13} - \theta^{14} \\ &+ \theta^{15} - \theta^{16} + \theta^{17} - \theta^{18}  + \theta^{19} - \theta^{20} + \theta^{21}  - \theta^{22}  + \theta^{23}   - \theta^{24}  + \theta^{25}  \end{aligned}$ & \SI{3.238150e-17}{}  \vspace{0.5cm} \\
		4 & $\begin{aligned} u_{1}(\bar{\theta})  =& 1.0 \theta - 1.0 \theta^{2} + 1.0 \theta^{3} - 1.0 \theta^{4} + 1.0 \theta^{5}  - 1.0 \theta^{6}  + 1.0 \theta^{7} - 1.0 \theta^{8} + 1.0 \theta^{9} - 1.0 \theta^{10} \\ &+ 1.0 \theta^{11}  - 1.0 \theta^{12} + 1.0 \theta^{13} - 1.0 \theta^{14} + 1.0 \theta^{15} - 1.0 \theta^{16} + 1.0 \theta^{17} - 1.0 \theta^{18}  \end{aligned}$ & \SI{4.876260e-13}{}  \vspace{0.5cm} \\
		5 & $\begin{aligned} u_{1}(\bar{\theta})  =& \theta  - \theta^{2} + \theta^{3}- \theta^{4} + \theta^{5}  - \theta^{6} + \theta^{7} - \theta^{8} + \theta^{9}  - \theta^{10} + \theta^{11} - \theta^{12}+ \theta^{13} - \theta^{14} + \theta^{15} \\ &- \theta^{16} + \theta^{17} - \theta^{18}  + \theta^{19} - \theta^{20} + \theta^{21}  - \theta^{22}  + \theta^{23}   - \theta^{24}   \end{aligned}$ & \SI{1.159374e-16}{}  \\
		\bottomrule  
	\end{tabular}
\end{table}
\begin{table}[H]
	\centering
	\caption{SR expansions for 5 samples ($\delta \gg 1$) for inputs $\{\delta, \delta^2, \delta^3\}$.}
	\label{tab:P1_Fit_Large-_HO}
	
	\begin{tabular}{c p{13cm} c}
		\toprule
		N & Best Approximation &       Fitness \\
		\midrule
		1 &  $\begin{aligned}  u_{1}(\bar{\eta}) =&  1.0 - 2.00 \eta + 1.80 \eta^{2} + 0.23 \eta^{3} - 4.71 \eta^{4} + 1.0 \eta^{5} - 3.03 \eta^{6}  - 1.0 \eta^{7} + 6.03 \eta^{8} \\ &+ 3.0 \eta^{9} - 2.0 \eta^{10} - 1.0 \eta^{12} - 1.00 \eta^{13}  \end{aligned}$ & \SI{1.73e-05}{}  \\
		2 &  $\begin{aligned} u_{1}(\bar{\eta}) =& 1.0 - 2.0 \eta + 2.0 \eta^{2} - 2 \eta^{3} + 2 \eta^{4} - 1.87\eta^{5} + 0.40\eta^{6} + 0.70\eta^{7} + 0.89\eta^{8}  \\ &+ 1.03\eta^{9}  + 0.30 \eta^{11}    \end{aligned}$ & \SI{1.01e-09}{}  \\
		3 &  $\begin{aligned}  u_{1}(\bar{\eta}) =&  1.0 - 2.0 \eta + 2.0 \eta^{2} - 2.0 \eta^{3} + 1.98 \eta^{4} - 1.38 \eta^{5} - 2.82 \eta^{6} + 0.62 \eta^{7} + 1.21 \eta^{9} \\ &+  0.61 \eta^{10}   \end{aligned}$ & \SI{7.96e-10}{}  \\
		4 &  $\begin{aligned} u_{1}(\bar{\eta}) =&  1.0 - 2.0 \eta + 2 \eta^{2} - 2.0 \eta^{3}+ 2.0 \eta^{4} - 2.0 \eta^{5}  + 2 \eta^{6}  - 2 \eta^{7} + 2.0 \eta^{8} - 2.0 \eta^{9} \\ &+ \eta^{10} +\eta^{11}   \end{aligned}$ & \SI{7.73e-15}{}  \\
		5 &  $\begin{aligned}  u_{1}(\bar{\eta}) =& 1.0 - 2.0 \eta + 2.0 \eta^{2} - 2.0 \eta^{3} + 2.0 \eta^{4}  - 2.0 \eta^{5} + 2.0 \eta^{6}  - 2.0 \eta^{7}  + 2.0 \eta^{8} - 2.0 \eta^{9}   \\ &+ 2.0 \eta^{10} - 2.0 \eta^{11} + 2.0 \eta^{12} - 2.0 \eta^{13} + 2.0 \eta^{14}   - 1.0 \eta^{15}   \end{aligned}$ & \SI{1.17e-16}{}  \\
		\bottomrule  
	\end{tabular}
\end{table}

\clearpage
\section*{Kelvin-Voigt Viscoelastic Solid} 
\begin{table}[H]
	\centering
	\caption{SR expansions for 5 samples ($\delta \ll 1$) .}
	\label{tab:P2_Fit_Small}
	\begin{tabular}{c p{13cm} c}
		\toprule
		N & Best Approximation &       Fitness \\
		\midrule
		1 &  $\begin{aligned} I(\delta) =  1.0 - \delta + 1.86 \delta^{2} - 3.03 \delta^{3}  - 0.24\delta^{4}  + 5.97\delta^{5}  +5.0 \delta^{6}  \end{aligned}$ & \SI{3.23e-05}{}  \vspace{0.5cm} \\
		2 &  $\begin{aligned} I(\delta) = 1.0 - 1.00 \delta + 1.78 \delta^{2}- 2.79 \delta^{3} 2.0 \delta^{4}  \end{aligned}$ & \SI{4.97e-05}{}  \vspace{0.5cm} \\
		3 &  $\begin{aligned} I(\delta) = 1.0- \delta + 1.96 \delta^{2} - 4.42\delta^{3} + 3.54\delta^{4} + 11.23\delta^{5}  - 0.27\delta^{6} + 0.04\delta^{7} +2.0 \delta^{8}    \end{aligned}$ & \SI{8.71e-06}{}  \vspace{0.5cm} \\
		4 &  $\begin{aligned} I(\delta) =  1.0 - 0.943 \delta + \delta^{2}  \end{aligned}$ & \SI{5.09e-04}{} \vspace{0.5cm} \\
		5 &  $\begin{aligned} I(\delta) =  1.0- 1.0 \delta + 1.95 \delta^{2} - 4.74\delta^{3} + 9.16\delta^{4} - 9.12\delta^{5} - 1.48\delta^{6} + 19.14\delta^{7} - 15.34\delta^{8}   \end{aligned}$ & \SI{6.01e-06}{}  \\
		\bottomrule  
	\end{tabular}
\end{table}

\begin{table}[H]
	\caption{SR expansions for 5 samples ($\delta \gg 1$) .}  
	\label{tab:P2_Fit_Large}
	\begin{tabular}{c p{13cm} c}
		\toprule
		N & Best Approximation &       Fitness \\
		\midrule
		1 &  $\begin{aligned} I(\eta) =& \eta (- \log{\eta} - 0.58) + \eta^{2} (0.45 - \log{\eta}) + 1.49 \eta^{3}  - 0.38 \eta^{4} - 0.67 \eta^{5}  + 0.58 \eta^{6}   \end{aligned}$ & \SI{2.46e-06}{}  \vspace{0.5cm} \\
		2 &  $\begin{aligned} I(\eta) =& \eta (- \log{\eta} - 0.58) + \eta^{2}  (0.42 - 1.0 \log{\eta}) + \eta^{3}  (0.01 \log{\eta}^{2} - 0.38 \log{\eta} + 0.79)  \\ &+ \eta^{4} (- 0.06 \log{\eta}^{2} + 0.09 \log{\eta}) + \eta^{5}  (0.01 \log{\eta}^{3} - 0.03 \log{\eta}^{2} - 0.01 \log{\eta}) \\ &+ \eta^{6} (- 0.01 \log{\eta}^{4} - 0.04 \log{\eta}^{3} + 0.02 \log{\eta}^{2}) + 0.02 \eta^{7} \log{\eta}^{2}  \end{aligned}$ & \SI{2.76e-09}{}  \vspace{0.5cm} \\
		3 &  $\begin{aligned} I(\eta) =& \eta (- \log{\eta} - 0.58) + \eta^{2}  (0.42 - \log{\eta}) + \eta^{3}  (0.36 - 0.53 \log{\eta})  +0.94\eta^{4}  \end{aligned}$ & \SI{1.91e-07}{}  \vspace{0.5cm} \\
		4 &  $\begin{aligned} I(\eta) =&  \eta (- 1.0 \log{\eta} - 0.58) + \eta^{2}  (1 - \log{\eta}) + \eta^{3}  (0.43 \log{\eta}^{3} - 0.25 \log{\eta}^{2} - 2.22 \log{\eta})  \\ &+ \eta^{4}  (0.17 \log{\eta}^{4} + 1.65 \log{\eta}^{3} - 2.79 \log{\eta}^{2} - 3.21 \log{\eta}) \\ &+ \eta^{5} (- 0.17 \log{\eta}^{5} 0.54 \log{\eta}^{4} + 1.91 \log{\eta}^{3} - 5.64 \log{\eta}^{2} + 0.21 \log{\eta})  \\ &+ \eta^{6} (- 1.02 \log{\eta}^{5} - 0.56 \log{\eta}^{4} + 2.02 \log{\eta}^{3} + 0.57 \log{\eta}^{2} + 1.34 \log{\eta}) \\ &+ \eta^{7} (- 1.88 \log{\eta}^{5} - 2.2 \log{\eta}^{4} + 3.6 \log{\eta}^{3} - 0.44 \log{\eta}) \\&+ \eta^{8} (- 1.0 \log{\eta}^{5} - 0.17 \log{\eta}^{4} + 4.13 \log{\eta}^{3} - 5.24 \log{\eta}^{2} + 1.0 \log{\eta})  \\ &+ \eta^{9}  (0.61 \log{\eta}^{3} + 4.41 \log{\eta}^{2} + 3.29 \log{\eta})    + \eta^{10}  (1.0 \log{\eta}^{3} + 0.9 \log{\eta}^{2} - 2.58 \log{\eta})   \end{aligned}$ & \SI{1.97e-05}{}  \\
	\end{tabular}
\end{table}
\begin{table}[H]
	\begin{tabular}{c p{13cm} c}
		\toprule
		N & Best Approximation &       Fitness \\
		\midrule
		5 &  $\begin{aligned} I(\eta) =& - 0.87 \eta \log{\eta} + \eta^{2} \log{\eta}^{2} (0.02 \log{\eta}^{3} - 0.04 \log{\eta}^{1} + 0.02) \\
			&+ \eta^{3} \log{\eta} (0.03 \log{\eta}^{5} + 0.12 \log{\eta}^{4} - 0.01 \log{\eta}^{3} - 0.3 \log{\eta}^{2} + 0.19\log{\eta}^{1} - 0.04 ) \\ 	
			&+ \eta^{4} \log{\eta} (0.01 \log{\eta}^{6} + 0.01 \log{\eta}^{5} + 0.11 \log{\eta}^{4} + 0.03 \log{\eta}^{3} - 0.51 \log{\eta}^{2}) \\ 
			&+ \eta^{4} \log{\eta} (0.48 \log{\eta}^{1} - 0.14 ) + \eta^{5} \log{\eta} (- 0.02 \log{\eta}^{8} + 0.01 \log{\eta}^{7} + 0.10 \log{\eta}^{6} ) \\ 
			&+ \eta^{5} \log{\eta} (+ 0.03 \log{\eta}^{5} + 0.34 \log{\eta}^{4} + 0.2 \log{\eta}^{3} - 1.59 \log{\eta}^{2} + 1.15 \log{\eta}^{1} - 0.21 ) \\ 
			&+ \eta^{6} (- 0.54 \log{\eta}^{5} + 1.16 \log{\eta}^{4} - 1.95 \log{\eta}^{3} + 1.83 \log{\eta}^{2} - 	0.61 \log{\eta}) \\ 
			&+ \eta^{6} (- 0.01 \log{\eta}^{10} - 0.06 \log{\eta}^{9} + 0.01 \log{\eta}^{8} + 0.37 \log{\eta}^{7} - 0.19 \log{\eta}^{6}) \\ 
			&+ \eta^{7} (- 2.52 \log{\eta}^{5} + 3.56 \log{\eta}^{4} - 2.89 \log{\eta}^{3} + 1.76 \log{\eta}^{2} - 0.30 \log{\eta}) \\ 
			&+ \eta^{7} (- 0.03 \log{\eta}^{10} - 0.19 \log{\eta}^{9} + 0.05 \log{\eta}^{8} + 1.18 \log{\eta}^{7} - 0.62 \log{\eta}^{6} ) \\ 
			&+ \eta^{8} (- 0.06 \log{\eta}^{10} - 0.42 \log{\eta}^{9} - 0.06 \log{\eta}^{8} + 2.79 \log{\eta}^{7} - 0.90\log{\eta}^{6}) \\ 
			&+ \eta^{8} ( - 6.74 \log{\eta}^{5} + 9.32 \log{\eta}^{4} - 5.06 \log{\eta}^{3} + 1.6 \log{\eta}^{2} - 0.46 \log{\eta}) \\ 
			&+ \eta^{9} (- 0.09 \log{\eta}^{10} - 0.66 \log{\eta}^{9} - 0.28 \log{\eta}^{8} + 4.8 \log{\eta}^{7} - 0.80 \log{\eta}^{6}) \\ 
			&+ \eta^{9} ( - 13.55 \log{\eta}^{5} + 18.17 \log{\eta}^{4} - 9.65 \log{\eta}^{3} + 2.24 \log{\eta}^{2} - 0.17 \log{\eta}) \\ 
			&+ \eta^{10} (- 0.11 \log{\eta}^{10} - 0.89 \log{\eta}^{9} - 0.52 \log{\eta}^{8} + 6.85 \log{\eta}^{7} - 0.30 \log{\eta}^{6}) \\ 
			&+ \eta^{10} (- 21.87 \log{\eta}^{5} + 28.7 \log{\eta}^{4} - 14.81 \log{\eta}^{3} + 3.14 \log{\eta}^{2} - 0.2 \log{\eta}) \\ 
			&+ \eta^{11} (- 27.65 \log{\eta}^{5} + 34.6 \log{\eta}^{4} - 17.26 \log{\eta}^{3} + 3.22 \log{\eta}^{2} - 0.07 \log{\eta}) \\ 
			&+ \eta^{11} (- 0.05 \log{\eta}^{10} - 0.75 \log{\eta}^{9} - 1.1 \log{\eta}^{8} + 6.92 \log{\eta}^{7} + 2.15 \log{\eta}^{6}) \\ 
			&+ \eta^{12} (- 0.25 \log{\eta}^{9} - 0.88 \log{\eta}^{8} + 3.95 \log{\eta}^{7} + 3.64 \log{\eta}^{6} - 24.62 \log{\eta}^{5}) \\
			&+ \eta^{12} ( 30.5 \log{\eta}^{4} - 14.11 \log{\eta}^{3} + 1.4 \log{\eta}^{2} + 0.36 \log{\eta})  \\
			&+ \eta^{13} (- 0.19 \log{\eta}^{8} + 1.28 \log{\eta}^{7} + 2.47 \log{\eta}^{6} - 14.48 \log{\eta}^{5}) \\ 
			&+ \eta^{13} (18.14 \log{\eta}^{4} - 7.53 \log{\eta}^{3} - 0.19 \log{\eta}^{2} + 0.49 \log{\eta}) \\ 
			&+ \eta^{14}  (0.27 \log{\eta}^{7} + 0.7598 \log{\eta}^{6} - 4.87 \log{\eta}^{5} + 6.86 \log{\eta}^{4} - 2.98 \log{\eta}^{3}) \\ 
			&+ \eta^{14}  ( - 0.56 \log{\eta}^{2} + 0.51 \log{\eta}) + \eta^{15}  (0.03 \log{\eta}^{6} - 0.66 \log{\eta}^{5} ) \\
			&+ \eta^{15}  (1.51 \log{\eta}^{4} - 0.8701 \log{\eta}^{3} - 0.29 \log{\eta}^{2} + 0.29 \log{\eta}) \\ 
			&+ \eta^{16} (- 0.04 \log{\eta}^{5} + 0.09 \log{\eta}^{4} - 0.05 \log{\eta}^{3} - 0.02\log{\eta}^{2} + 0.02 \log{\eta}) 
		\end{aligned}$ & \SI{4.74e-06}{}  \vspace{0.5cm} \\
		\bottomrule  
	\end{tabular}
\end{table}

\clearpage
\section*{Elastic Bending Wave}

\begin{table}[H]
	\centering
	\caption{SR expansions for 5 samples.}
	\label{tab:P3_Fit_Small}
	\begin{tabular}{c p{13cm} c}
		\toprule
		N & Best Approximation &       Fitness \\
		\midrule
		1 &  $\begin{aligned} K^{4} = - 0.01 \Omega + 1.01 \Omega^{2} + 1.45 \Omega^{3}  + 0.6 \Omega^{4} + 0.16 \Omega^{5}   \end{aligned}$ & \SI{2.14e-4}{}  \vspace{0.5cm} \\
		2 &  $\begin{aligned} K^{4} =& 0.97 \Omega^{2} + 1.53 \Omega^{3}  + 0.57 \Omega^{4} + 0.07 \Omega^{5}  + 0.08 \Omega^{6} + 0.03 \Omega^{7}  - 0.03 \Omega^{8}  - 0.03 \Omega^{9}   \\ &- 0.01 \Omega^{10} + 0.01 \Omega^{11} + 0.01 \Omega^{12}  \end{aligned}$ & \SI{4.71e-05}{}  \vspace{0.5cm} \\
		3 &  $\begin{aligned} K^{4} =  1.0 \Omega^{2} + 1.33 \Omega^{3} + 0.97 \Omega^{4}  - 0.21 \Omega^{5}  +0.12 \Omega^{6} \end{aligned}$ & \SI{1.97e-05}{}  \vspace{0.5cm} \\
		4 &  $\begin{aligned} K^{4} = 0.95 \Omega^{2} + 1.56 \Omega^{3}  + 0.54 \Omega^{4} + 0.15 \Omega^{5} + 0.03 \Omega^{6}  - 0.02 \Omega^{7}  \end{aligned}$ & \SI{7.53e-05}{}  \vspace{0.5cm} \\
		5 &  $\begin{aligned} K^{4} = 0.99 \Omega^{2} + 1.39 \Omega^{3}+ 0.81 \Omega^{4} - 0.03 \Omega^{5}  + 0.05 \Omega^{6}   \end{aligned}$ & \SI{3.6707e-05}{}  \vspace{0.5cm}  \\
		\bottomrule  
	\end{tabular}
\end{table}

\bibliography{bibliography.bib} 

\end{document}